\documentstyle[prl,floats,aps]{revtex}

\begin{document}

\draft

\twocolumn[\hsize\textwidth\columnwidth\hsize\csname @twocolumnfalse\endcsname

\title{A simple construction of initial data for multiple black holes}
\author{Steven Brandt and Bernd Br\"ugmann}
\address{Max-Planck-Institut f\"ur Gravitationsphysik
 \\ Schlaatzweg 1, 14473 Potsdam, Germany 
 \\ brandt@aei-potsdam.mpg.de, bruegman@aei-potsdam.mpg.de}
\date{February 11, 1997}
\maketitle

\begin{abstract}
We consider the initial data problem for several black holes in vacuum
with arbitrary momenta and spins on a three space with punctures.  We
compactify the internal asymptotically flat regions to obtain a
computational domain without inner boundaries.  When treated
numerically, this leads to a significant simplification over the
conventional approach which is based on throats and isometry
conditions.  In this new setting it is possible to obtain existence and
uniqueness of solutions to the Hamiltonian constraint.
\end{abstract}
\pacs{04.25.Dm, 95.30.Sf, 97.60.Lf; gr-qc/9703066; to appear in Phys.\ Rev.\ Lett.}

\vskip2pc]

\narrowtext


Binary black hole spacetimes are one of the great challenges for
numerical general relativity, even if no matter sources are present.
Here we consider the problem of finding initial data for several black
holes in vacuum with arbitrary momenta and spins. In general
relativity, initial data on a hypersurface cannot be specified freely,
because the Einstein equations give rise to four equations, three momentum
constraints and the Hamiltonian constraint, that the initial data has 
to satisfy. The purpose of this note is to introduce a
novel approach which is significantly simpler than the conventional
method based on throats and conformal imaging.

In all that follows we will assume vacuum, that the metric is
conformally flat, and that the extrinsic curvature is tracefree. 
A convenient form of the constraints of general relativity
can be obtained by rescaling the physical three-metric $g^{ph}_{ab}$
and its extrinsic curvature $K^{ph}_{ab}$ by a conformal factor
$\psi$,
\begin{equation}
  g^{ph}_{ab} = \psi^4 g_{ab}, \quad
  K^{ph}_{ab} = \psi^{-2} K_{ab}.
\end{equation}
The momentum constraint becomes
\begin{equation}
  \nabla_a K^{ab} = 0,
\label{diffeo}
\end{equation}
and the Hamiltonian constraint becomes an elliptic equation for the scalar 
field $\psi$,
\begin{equation}
  \Delta \psi + \frac{1}{8} K^{ab}K_{ab} \psi^{-7} = 0,
\label{hamil}
\end{equation}
where the covariant derivatives are defined by the flat metric
$g_{ab}$, which is also used to raise and lower indices (see
\cite{Li44,Yo79}).  

In order to obtain black hole vacuum data, one has to introduce a
non-trivial topology. The first calculations were performed by
Einstein and Rosen \cite{EiRo33} in their work on point particles in
general relativity. Various constructions for black holes based on
Einstein-Rosen bridges and ``wormholes'' were given in e.g.\
\cite{Wh55,MiWh57,Mi63,BrLi63}.  The spatial slice typically consists
of two or more copies of $R^3$ with several spheres removed and
identifications of the various spherical inner boundaries. In this way
several asymptotically flat regions are obtained that are connected by
bridges or ``throats''.

The simplest example derives from the Schwarzschild spacetime in quasi
isotropic coordinates.  Considered as a problem on $R^3$ minus the
point $r = 0$, the constraint equations (\ref{diffeo}) and
(\ref{hamil}) are solved by
\begin{equation}
   \psi = 1 + \frac{m}{2r}, \quad K_{ab} = 0,
\label{SS}
\end{equation}
where $m$ is the mass and $r$ the isotropic radius.  To make contact
with the throat picture, recall that there exists an isometry given by
$r \leftrightarrow m^2/(4r)$ which leaves the coordinate sphere $r =
m/2$ invariant and which maps the entire exterior asymptotically flat
space into that sphere. Consequently, there exists a second
asymptotically flat region near $r = 0$. Equivalently,
one can represent this solution to the constraints on a space
consisting of two copies of $R^3$ with a sphere excised and
appropriate identification at the spheres.

For $N$ black holes and non-vanishing extrinsic curvature, York and
others \cite{Yo89} have developed a sophisticated method to solve the
constraints for two asymptotically flat spaces that are connected by
as many throats (i.e.\ excised spheres) as there are black holes, and
that are isometric copies of each other.
Note that there are explicit solutions to the momentum
constraint (\ref{diffeo}) that characterize a single black hole with
given momentum $P^a$, and spin $S^a$. For example,
\begin{eqnarray}
   K^{ab}_{PS} &=& 
   \frac{3}{2r^2} (P^a n^b + P^b n^a - (g^{ab} - n^a n^b) P^c n_c) 
\nonumber\\
   && + \frac{3}{r^3} (\epsilon^{acd} S_c n_d n^b 
                      +\epsilon^{bcd} S_c n_d n^a),
\label{KPJ}
\end{eqnarray}
where $n^a$ is the radial normal vector.
Since the conformal metric is flat, we will use in what follows
either standard spherical or Cartesian coordinates, i.e.
$r = (x^2+y^2+z^2)^{1/2}$ and $n^a = x^a/r$. 

By the method of images it is possible to obtain an infinite series
based on (\ref{KPJ}) for $K^{ab}$ which solves the momentum constraint
and satisfies an isometry condition at any number of spheres
\cite{Mi63,Yo89}. Given such a solution, what remains to be done is to
solve the Hamiltonian constraint (\ref{hamil}), which is an elliptic
equation on $R^3$ minus several spheres, with the inner boundary given
by the isometry and the outer boundary determined by asymptotic
flatness.

Three independent numerical implementations of the above scheme have
been given and compared in the definitive paper on three dimensional
black hole initial data by Cook, Choptuik, Dubal, Klasky, Matzner, and
Oliveira \cite{CCDKMO93}. For the Hamiltonian constraint, they
consider a non-linear block full approximation storage multigrid
scheme for Cadez coordinates, a successive over-relaxation
scheme in Cartesian coordinates, and a multiquadratics approach. All
these approaches are greatly complicated by the presence of the inner
spherical boundaries and the isometry condition.


Let us return to the Schwarzschild solution of the constraints on a
``punctured'' $R^3$. As noted by Misner and Wheeler \cite{MiWh57}, 
and studied in detail by Brill and Lindquist \cite{BrLi63}, the
Schwarzschild solution to the constraints generalizes trivially to $N$
black holes for time symmetry,
\begin{equation}
\psi = 1 + \sum_{i=1}^{N} 
\frac{m_{(i)}}{2 \left|\vec{r}-\vec{r}_{(i)}\right|},
\quad
K_{ab}=0
\label{BL}
\end{equation}
where $m_{(i)}$ characterizes the mass of the $i$-th black hole 
(i.e.\ the ADM mass is $\sum m_{(i)}$) and $\vec{r}_{(i)}$ is the
location of the $i$-th black hole.  For regularity of the conformal
factor, the MWBL solution is considered on a single $R^3$ with the
points $\vec{r}=\vec{r}_{(i)}$ removed. We refer to the
$\vec{r}_{(i)}$ as punctures.
The isometry present in the Schwarzschild solution is lost, although
there still exist minimal surfaces characterizing the throats
\cite{BrLi63}. 

Let us now discuss the method that we propose to find data for
multiple black holes with arbitrary boosts and spins.  
The idea is to compactify the internal asymptotically flat regions
in order to obtain a simple domain of integration. Such
compactification brings up issues of regularity of the fields
(e.g. \cite{Fr88,BeMu94}), which we address below.

As before, we
consider vacuum spacetimes, the metric is conformally flat, and the
extrinsic curvature is tracefree.  As the spatial slice we choose a
single $R^3$ with $N$ punctures as in the MWBL data. First, we solve
the momentum constraint (\ref{diffeo}) by setting
\begin{equation}
  K^{ab} = \sum_{i = 1}^{N} K_{PS(i)}^{ab},
\label{Ksol}
\end{equation}
where each term is defined by (\ref{KPJ}) with its own origin
$\vec{r}_{(i)}$, momentum $\vec{P}_{(i)}$, and spin
$\vec{S}_{(i)}$. These parameters correspond to the ADM quantities
in the limit that the separation of the holes is very large.  The
equation (\ref{Ksol}) defines the solution to the momentum
constraint that we actually use, and is not just the starting point
for the method of images that is usually invoked to distort $K_{ab}$
to obtain an isometric solution (cmp.\ \cite{Th87}, where the same
simplification arises for a trapped surface boundary condition at the
inner boundary).

Given $K_{ab}$ as defined in (\ref{Ksol}), we proceed to solve the
Hamiltonian constraint, (\ref{hamil}).  We rewrite the conformal
factor in terms of functions $\alpha$ and $u$ given by
\begin{equation}
  \psi = \frac{1}{\alpha} + u, \quad 
  \frac{1}{\alpha} = \sum_{i=1}^{N} 
\frac{m_{(i)}}{2 \left|\vec{r}-\vec{r}_{(i)}\right|}.
\label{ourpsi}
\end{equation}
On the punctured $R^3$, the Laplacian of
$1/\alpha$ is zero, so that the Hamiltonian constraint equation
becomes (cmp.\ \cite{BeMu94} for a single asymptotic region and vanishing
linear momentum)
\begin{eqnarray}
  \Delta u + \beta (1+\alpha u)^{-7} = 0,
\label{keyeq}
\\
  \beta = \frac{1}{8} \alpha^{7} K^{ab} K_{ab}.
\end{eqnarray}
To complete the definition of the problem, we have to specify boundary
conditions for $u$. For asymptotic flatness at infinity we require
$u - 1 = O(r^{-1})$ for large distances to the punctures. 

The key question that remains is what condition we want to impose on
$u$ close to the punctures. As it turns out, to build in
asymptotically flat regions as are present in the MWBL data near the
punctures, {\em it suffices to solve (\ref{keyeq}) everywhere
on $R^3$ without any points excised.} This completes the statement of
our proposal.

Let us discuss existence and uniqueness of solutions to the modified
Hamiltonian constraint equation (\ref{keyeq}) on $R^3$. Since in this
case the topology is trivial, we can show existence and uniqueness of
a $C^2$ solution by repeating the proof given in \cite{Ca79} for the
conventional Hamiltonian constraint equation on $R^3$. 
By definition,
both $\alpha$ and $\beta$ are proportional to 
$|\vec{r}-\vec{r}_{(i)}|$ near the former punctures and are therefore
$C^0$, despite the fact that $K_{ab} K^{ab}$ goes as 
$|\vec{r}-\vec{r}_{(i)}|^{-6}$ at these
points. From the maximum principle and the outer boundary condition we
obtain that there exists at most one $C^2$ solution, and in
particular, that $u \geq 1$ and $1+\alpha u \geq 1$.  Referring to
\cite{Ca79} for the definition of weighted Sobolev spaces, for the
existence part of the proof we require that $u \in M^p_{s,\delta}$ and
$\beta (1+\alpha u)^{-7} \in M^p_{s-2,\delta+2}$ with $p > 3$ for the
norm and $s \geq 3$ characterizing differentiability. The fall-off of
(\ref{keyeq}) is the same as in the standard case. Interior points
like the $\vec{r}_{(i)}$ do not affect the weight $\delta$. Due to the
loss of differentiability at the punctures in $\beta$, we find that in
our case $p>3$ and $s = 3$.  (In comparison, \cite{ChMu81} uses $p=2$
but requires $s\geq 4$). The
proof proceeds as in \cite{Ca79} with minor changes in the algebra and
due care whenever it matters that $\alpha$ and $\beta$ may vanish.
So, although $u$ is only $C^2$ at the $\vec{r}_{(i)}$,
there exists a unique solution for the conformal factor $\psi = u +
1/\alpha$ on the punctured $R^3$ determined by our proposal for $u$ on
the unpuctured $R^3$.


Given a solution $u$, we can demonstrate that each puncture represents
the ``point at infinity'' for another asymptotically flat spacetime
(cmp.\ \cite{BrLi63}). Hence, solving (\ref{keyeq}) on $R^3$ involves
a particular compactification of $N$ out of $N+1$ asymptotically flat
regions, one of which is distinguished by our choice of $K_{ab}$, see
below. We perform a coordinate inversion through a sphere near the
i-th puncture, $ \bar{r} = a^2/r $, under which the metric transforms
as
\begin{equation}
ds^2 = \psi^4 \left(dr^2 + r^2 d\Omega^2\right) = 
  \bar{\psi}^4 \left(d\bar{r}^2+\bar{r}^2 d\Omega^2\right),
\end{equation}
with
$
\bar{\psi} = \psi \, r / a
$ (which is not an isometry).
Setting $a = m_{(i)}/2$ where $m_{(i)}$ is the bare mass of the
puncture we are considering, we obtain for our choice of $\psi$,
(\ref{ourpsi}), that
\begin{eqnarray}
  \bar\psi &=& 1 + \frac{\bar m_{(i)}}{2\bar r} + O(\frac{1}{\bar r^2}),
\\
  \bar m_{(i)} &=& m_{(i)} \left( u(\vec{r}_{(i)}) + 
  \sum_{j\neq i} \frac{m_{(j)}}
  {2 \left|\vec{r}_{(i)}-\vec{r}_{(j)}\right|} \right).
\end{eqnarray}
Therefore, the metric becomes flat as we approach the punctures.  

To show asymptotic flatness at the punctures, it remains to be shown
that $\bar K_{ab} = O(\bar r^{-2})$. In the physical variables, we
just have the coordinate transformation 
$\bar{r} = a^2/r$, 
$\Lambda^b_c = \partial x^b / \partial\bar x^c = (a^2/\bar r^2) L^b_c$ 
where
$L^b_c = \delta^b_c - 2 n^b n_c$. In the unphysical variables, we
also have to take into account the transformation of $\psi$. With   
$K_{ab} = \bar\psi^{2} \bar K_{ab}^{ph}$ and
$\bar K_{ab}^{ph} = (\Lambda K^{ph})_{ab}$, we obtain
\begin{equation}
  \bar K_{ab} = \left(\frac{a}{\bar r}\right)^6 (LK)_{ab}.
\end{equation}  
Therefore, for our choice of $K_{ab}$, (\ref{KPJ}), the momentum term
of order $r^{-2}$ is mapped to order $\bar r^{-4}$ while the spin term
of order $r^{-3}$ is mapped to order $\bar r^{-3}$.  This observation
extends to multiple hole data with $K_{ab}$ defined in (\ref{Ksol}).
Hence the black hole puncture data is asymptotically flat at the
punctures.

In fact, we also learn that the black hole seen from the region near
the puncture appears to be unboosted, because the boost term is obtained
from the $r^{-4}$ term in the untransformed space. This term is
always zero in our space by construction.  In the conventional method,
imposing the isometry leads to terms which go as $r^{-4}$ in $K_{ab}$.
These terms may not be included in our method (without modification),
because $\beta$ would not be regular at the punctures if they were.

Since the initial data is asymptotic to Schwarzschild data near the
punctures, and since small spherical surfaces centered at the puncture
of Schwarzschild data are outer-trapped surfaces (with outside
referring to the asymptotic region away from the punctures), this
indicates that the region near the punctures are inside black holes.
As in \cite{BrLi63}, several punctures may be hidden behind a common
horizon, which we confirmed numerically but will not discuss here.


We now come to the numerical implementation of our proposal.  
We found that a full approximation storage multigrid method
built around a non-linear Gauss-Seidel relaxation scheme
\cite{Br82} performs very well for (\ref{keyeq}) on a 
finite Cartesian grid with a Robin boundary condition.  
Note that the standard representation of $\Delta u$ by centered finite
differences is only first order at the punctures. One could use a
different prescription at the puncture points, but in numerical tests
we found that the correspondingly lower rate of convergence is
contained in a surprisingly small neighborhood of the puncture points,
whether those are part of the grid or not.

The multigrid program is part of BAM, a bifunctional adaptive mesh
package for elliptic and hyperbolic problems in three-dimensional
numerical relativity (see \cite{Br96} for the hyperbolic part; we do
not use adaptivity here). Since in general the Hamiltonian constraint
has to be solved numerically, it is important to provide at least one
efficient numerical implementation.  There is no reason to believe
that the multigrid method is the only good method. However, when
available, multigrid methods have proven to be among the best
performers. We consider the absence of irregular boundaries in our
method to be a valuable feature since it makes a straightforward
multigrid implementation possible.


As a test, let us compute the correction to the conformal factor for
two equal mass black holes boosted toward one another when we keep
only terms of small $P$ (the momentum of each hole) and $L$ (the
coordinate positions of the holes along the z-axis being $\pm L$).
Consider $\beta$ to be of order $\epsilon$, and consider the MWBL
solution to be the zeroth order solution to the Hamiltonian
constraint, $u=u_{(0)}+\epsilon u_{(1)}= 1+\epsilon u_{(1)}$.  We now
have
\begin{equation}
\epsilon \Delta u_{(1)} = - \beta (1 + \alpha)^{-7} .
\end{equation}
The ADM mass to first order in $\epsilon$ is 
\begin{eqnarray}
M_{ADM} &=& -\frac{1}{2 \pi} \oint_{r=\infty} \!\!\!\! 
\nabla_a \left(\frac{1}{\alpha}\right) dS^a+
\frac{1}{2 \pi} \int \beta (1 + \alpha)^{-7} dV
\nonumber
\\
 = &2m& + m \left(\frac{P}{m}\right)^2 \left[
 \frac{11}{50}\left(\frac{L}{m}\right)^2
-\frac{24}{35}\left(\frac{L}{m}\right)^4 \right].\label{boo}
\end{eqnarray}
This equation predicts the correction to the ADM mass
resulting from the full non-linear solve to within about
$5\%$
when $P/m \leq 1.0$ and $L/m \leq 0.2$ and the grid is a cube whose
sides have a length of about $16 m$ and are resolved with $67$ zones.

This calculation provides one test for the correctness of the solution
of the data set, at least for the ADM mass of the spacetime for
perturbative cases.  This is valuable, because the code cannot be
tested for the case when $K_{ab}=0$ since then the solution is
trivially the MWBL solution.

\begin{table}
\begin{tabular}{rrrrrr}
\multicolumn{6}{c}{A2B8, box of size $32$}
\\\hline
\multicolumn{6}{c}{
\begin{tabular}{lclclcl}
$m_{(1)}$ &=& 2 &&
$m_{(2)}$ &=& 1
\\
$\vec{r}_{(1)}$ &=& $(0, 0, 4)$ &&
$\vec{r}_{(2)}$ &=& $(0, 0, -4)$
\\
$\vec{P}_{(1)}$ &=& $(15, 0, 0)$ &&
$\vec{P}_{(2)}$ &=& $(-15, 0, 0)$
\\
$\vec{S}_{(1)}$ &=& $(-20, 20, 0)$ &&
$\vec{S}_{(2)}$ &=& $(0, 20, 20)$
\end{tabular}
}
\\\hline
\multicolumn{1}{c}{$h$} & 
\multicolumn{1}{c}{$n$} & 
\multicolumn{1}{c}{$\|d\|_2$} & 
\multicolumn{1}{c}{$\|\tau\|_2$} & 
\multicolumn{1}{c}{$\sigma$} & 
\multicolumn{1}{c}{$M_{ADM}$} 
\\\hline
 4.000& $11^3$  & 1.159e-08 & 3.813e-03 & 4.029 & 9.209
\\
 2.000& $19^3$  & 5.350e-08 & 1.984e-03 & -2.119& 10.097
\\
 1.000& $35^3$  & 2.141e-07 & 5.400e-03 & 0.838 & 10.601
\\
 0.500& $67^3$  & 2.582e-07 & 1.216e-02 & 1.121 & 10.833
\\
 0.250& $131^3$ & 1.552e-07 & 2.222e-02 & 1.430 & 10.917
\\
 0.125& $259^3$ & 6.629e-08 & 3.383e-02 & 1.712 & 10.942
\end{tabular}

\caption{Results for the A2B8 data set of [9]. All units 
are in terms of $m_{(2)}$, $h$ is the grid spacing, $n$ the number of
grid points, $d$ is the residual, $\tau$ the truncation error,
$\sigma$ the convergence rate, and 
$M_{ADM}$ is the ADM mass of the box.}

\label{table}
\end{table}

As a general 3d strong field example we present in Tab.\ \ref{table}
data comparable to the A2B8 data set of \cite{CCDKMO93}. Note that
the data sets produced by the two methods are not identically the
same. They are equally generic, but each has a different gravitational
wave content.

There are two
punctures with different mass parameters of order one, and with
comparatively large values for boosts and spins. The results are
obtained for a single box centered around the origin for various grid
spacings. Each run was performed for the same multigrid parameters
such that the values for the residua reflect the efficiency with which
the discretized equations are solved at different resolutions.  We
chose to make the residua much smaller than the
truncation error estimate, 
which turns out to
be in a range compatible with the still rather large grid spacing. The
convergence rates $\sigma$ are based on $u$ at three resolutions. 
As expected, we found in this and other examples that
the convergence rate at the puncture points was one order less than
elsewhere, but we also found that convergence was only affected very
close to the punctures.  As the resolution increases, the mass
estimate converges with a rate comparable to $\sigma$.


To summarize, we have introduced a new setup for the initial data
problem for multiple black holes with arbitrary mass, momentum, and
spin in three dimensions.  Comparing with the wormhole constructions
of black holes, the key difference is how the various asymptotically
flat regions are defined on a single copy of $R^3$. Instead of using
an isometry condition at interior spheres, we remove the inner
boundary by compactification and look for a solution to a modified
Hamiltonian constraint equation on $R^3$, for which we have existence
and uniqueness.  A corresponding proof should be possible for the
isometric wormhole data, although to our knowledge, such a proof as
not been given yet.

There are two obvious numerical advantages compared to the wormhole
approach.  First note that the solution $K^{ab}$ to the momentum
constraint is all we need, i.e. since we do not impose the isometry
condition there is no need for the method of images. Second, in the
solution of the Hamiltonian constraint we avoid the
numerical complications due to an inner boundary. 
Compared to the standard Cartesian (Cadez) method, one (two)
levels of sophistication less are required for an implementation,
which should make general black hole initial data more widely
available.

Finally, note that black hole puncture data has already been
successfully evolved in the case of 3d Schwarzschild initial data
\cite{ACMSST95,Br96}, which is possible even when the punctures
are part of the numerical grid since there are no physical
singularities on the initial slice. For a long term evolution of black
holes, one may start with data obtained by either the throat
or puncture construction, and then cut out the interior regions at the
apparent horizon to avoid the physical singularities on future slices
\cite{Th87,SeSu92}.

It is a pleasure to thank Helmut Friedrich, Alan Rendall, Bernd
Schmidt, Ed Seidel, and Paul Walker for helpful discussions. The
computations were performed at the AEI in Potsdam.

\newcommand{\bb}{B. Br\"ugmann}
\newcommand{\bib}[1]{\bibitem{#1}}
\newcommand{\EM}{}
\newcommand{\apny}[1]{{\EM Ann.\ Phys.\ (N.Y.) }{\bf #1}}
\newcommand{\cjm}[1]{{\EM Canadian\ J.\ Math.\ }{\bf #1}}
\newcommand{\cmp}[1]{{\EM Commun.\ Math.\ Phys.\ }{\bf #1}}
\newcommand{\cqg}[1]{{\EM Class.\ Quan.\ Grav.\ }{\bf #1}}
\newcommand{\grg}[1]{{\EM Gen.\ Rel.\ Grav.\ }{\bf #1}}
\newcommand{\jgp}[1]{{\EM J. Geom.\ Phys.\ }{\bf #1}}
\newcommand{\ijmp}[1]{{\EM Int.\ J. Mod.\ Phys.\ }{\bf #1}}
\newcommand{\JCP}[1]{{\EM J. Comp.\ Phys.\ }{\bf #1}}
\newcommand{\jmp}[1]{{\EM J. Math.\ Phys.\ }{\bf #1}}
\newcommand{\mpl}[1]{{\EM Mod.\ Phys.\ Lett.\ }{\bf #1}}
\newcommand{\np}[1]{{\EM Nucl.\ Phys.\ }{\bf #1}}
\newcommand{\PL}[1]{{\EM Phys.\ Lett.\ }{\bf #1}}
\newcommand{\pr}[1]{{\EM Phys.\ Rev.\ }{\bf #1}}
\newcommand{\PRL}[1]{{\EM Phys.\ Rev.\ Lett.\ }{\bf #1}}

\end{document}